\documentclass[aps,preprint,showpacs,preprintnumbers,amsmath,amssymb,floatfix]{revtex4}
\usepackage[german,english]{babel}
\usepackage{bm,amsmath,amsfonts}
\usepackage{amssymb}
\usepackage{graphicx}
\usepackage{xcolor}

\def\beq{\begin{equation}}
\def\eeq{\end{equation}}
\def\beqn{\begin{eqnarray}}
\def\eeqn{\end{eqnarray}}

\begin{document}

\title{Death of soliton trains in attractive Bose-Einstein condensates} 

\author{Alexej I. Streltsov$^{1}$\footnote{E-mail: Alexej.Streltsov@pci.uni-heidelberg.de}, 
Ofir. E. Alon$^{1,2}$,
and Lorenz S. Cederbaum$^{1}$\footnote{E-mail: Lorenz.Cederbaum@pci.uni-heidelberg.de}}

\affiliation{$^1$ Theoretische Chemie, Physikalisch-Chemisches Institut, Universit\"at Heidelberg,\\
Im Neuenheimer Feld 229, D-69120 Heidelberg, Germany}

\affiliation{$^2$ Department of Physics, University of Haifa at Oranim, Tivon 36006, Israel}

\begin{abstract}
Experiments on ultra-cold attractive Bose-Einstein Condensates (BECs) have demonstrated that
at low dimensions atomic clouds can form localized objects, 
propagating for long times without significant changes in their shapes 
and attributed to bright matter-wave solitons,
which are coherent objects.
We consider the dynamics of bright soliton trains 
from the perspective of many-boson physics.
The fate of matter-wave soliton trains
is actually to quickly loose their coherence and become
macroscopically fragmented BECs. 
The death of the coherent 
matter-wave soliton trains gives 
birth to fragmented objects,
whose quantum properties and
experimental signatures differ 
substantially from what is 
currently assumed.
\end{abstract}

\maketitle

Solitons are common wave-packet phenomena in many areas of science and engineering \cite{sol1,sol2}. 
Such localized structures can form when
a wave-packet's dispersion is compensated by 
self-focusing (attractive) forces.
In this context, 
the physics of low-dimensional attractive 
Bose-Einstein condensates (BECs)
has attracted much attention \cite{solth1,solth2,exp1,exp2,attrac1,attrac2,attrac3,attrac4}.
The similarity between the Gross-Pitaevskii (GP)
equation \cite{GP1,GP2} commonly used
to describe these ultra-cold quantum systems
and the non-linear Schr\"odinger equation used in optics \cite{sol2} 
to characterize self-focusing of light 
has encouraged the transfer
of ideas, phenomena,
and understanding from
optics to ultra-cold atomic physics.
Particularly,
bright matter-wave solitons have been predicted
to occur in low-dimensional attractive Bose gases \cite{solth1,solth2},
stimulating the recent experiments \cite{exp1,exp2}.

Very often,
along with a single soliton,
multi-hump matter-wave structures  
propagating for long times without significant changes in their shapes 
have been observed.
Since the GP equation
supports localized multi-hump solutions as excited states,
the experimentally observed multi-hump
structures have been attributed to soliton trains,
which are coherent objects.
This is the current widely accepted interpretation in the literature.
Nowadays the GP equation (i.e., the non-linear Schr\"odinger equation)
is the cornerstone of studying soliton trains in attractive BECs.

It is important to remember, however,
that ultra-cold attractive BECs are quantum many-particle systems,
governed by the many-boson Schr\"odinger equation. 
Hence, we consider in this work the dynamics of bright soliton trains 
from the perspective of many-boson physics.
As we shall show the fate of matter-wave soliton trains
is to quickly loose their coherence and become
macroscopically fragmented BECs \cite{frag1,frag2,frag3}. 
The death of the coherent 
matter-wave soliton trains gives 
birth to fragmented objects,
whose quantum properties and
experimental signatures differ 
substantially from what is 
currently assumed.

Our starting point is the GP dynamics 
$\left[-\frac{1}{2} \frac{\partial^2}{\partial x^2} + \lambda_0(N-1) |\psi(x,t)|^2\right] \psi(x,t) = i\frac{\partial \psi(x,t)}{\partial t}$
of a soliton train with two humps (briefly, two-hump soliton).
We consider in this work
$N=2000$ bosons in one dimension 
and attractive inter-particle 
interaction of strength $\lambda_0 = - 0.002$.
For convenience, we use dimensionless quantities
which are readily arrived at by dividing
the dimension-full Hamiltonian
by $\frac{\hbar^2}{m L^2}$,
where $m$ is the mass of a boson 
and $L$ is a length scale.
The corresponding dimension-full quantities are within range of current 
experimental setups and discussed below.
As an initial profile (shape) of the two-hump soliton
we take the commonly used linear combination of sech-shaped functions:
$$
\psi(x,0) = {\cal N} \left\{ \mathrm {sech} [\gamma(x-x_0)] \pm \mathrm {sech} [\gamma(x+x_0)] \right\},
$$
where $\cal N$ is the normalization factor.
The "+" refers to 
a $0$-phase symmetric and the "-"
to a $\pi$-phase anti-symmetric soliton train.
The parameter $\gamma$ is inversely proportional 
to the width of the humps.
By assuming a large separation between humps, 
one gets by minimizing the GP energy functional the optimal value of 
$\gamma = |\lambda_0| (N-1)/4 \approx 1.0$.

The time-dependent 
density $\rho(x,t)$
obtained by solving
the time-dependent
GP equation for different two-hump solitons 
is plotted in Fig.~1.
In the left upper panel, Fig.~1a, 
we plot the dynamics for the $0$-phase symmetric soliton train, 
initially located at $x_0=\pm 4.0$. 
The system reveals the well known mean-field oscillatory dynamics:
the humps attract each other, collide at $t \approx 42$, 
then split and return to the initial separation and start a new cycle of oscillation.
The left middle part, Fig.~1b,
shows the dynamics for the $\pi$-phase anti-symmetric soliton clouds 
placed initially at a larger separation of $x_0=\pm 6.0$.
At such a separation the clouds very slightly repel each other, 
indicating that the inter-hump forces strongly depend on the initial separation.
In the left lower panel, Fig.~1c, 
we depict the
dynamics for a $0$-phase asymmetric soliton train,
formed as the combination of the sech-shaped functions of slightly 
different  widths $\gamma=1.1$ (left cloud) and $\gamma=0.9$ (right cloud),
placed at $x_0=\pm 4.0$.
A similar dynamics to Fig.~1a is seen.
The GP theory describes the following characteristics of two-hump solitons:
(i) the relative phase between humps defines whether the inter-hump interaction is attractive or repulsive,
and 
(ii) the magnitude of the inter-hump interaction depends on the separation between the humps.

We remind the reader that GP theory is a mean-field approximation to
the quantum many-boson problem which assumes 
all bosons to occupy one and the same quantum state
throughout the system's evolution in time.
What happens if we relax the constraint of the attractive system to occupy a single quantum state? 
In other words, 
we allow 
the system to choose according to the variational 
principle its evolution in time by solving now the 
time-dependent many-boson Schr\"odinger equation 
$i \frac{\partial \Psi(t)}{\partial t} = \hat H \Psi(t)$,
with
$\hat H(x_1,\ldots,x_N) = \sum_{j=1}^N - \frac{1}{2} \frac{\partial^2}{\partial x_j^2} + 
\sum_{j < k}^N \lambda_0 \delta(x_j-x_k)$.
It is possible nowadays to go significantly beyond
mean-field and compute the dynamics of 
BECs on the many-body level,
by solving the time-dependent many-boson Schr\"odinger equation
with the multiconfigurational time-dependent Hartree method for bosons,
see the literature for details \cite{MCTDHB_lett,MCTDHB_pap}
and applications \cite{Grond,BJJ_us}.
Is the dynamics of soliton trains described above 
to change in a dramatic manner?

We take as initial conditions 
the same $0$-phase symmetric and $\pi$-phase anti-symmetric 
soliton trains as in the above 
GP studies and propagate them now not on the mean-field but on the many-body level.
The time-dependent density $\rho(x,t)$ 
obtained at the many-body level is plotted in the right panels of Fig.~1.
The first observation is that the initially-localized wave-packets 
remain localized at the many-body level of the description as well.
Hence, localization is indeed a characteristic dynamical feature of 
attractive bosonic clouds in one dimension,
as predicted at the GP level and confirmed here 
at the many-body level.
The many-body results, however, reveal different dynamics than the mean-field ones,
compare the left and right panels of Fig.~1.
In the right upper panel, Fig.~1d,
we depict the many-body dynamics of the initially-prepared $0$-phase symmetric wave-packet.
The two separated humps start to attract each other, 
similarly to the mean-field dynamics,
but
do not come close enough to collide,
as it is in the GP case, Fig.~1a. 
Instead of the full collision the sub-clouds reach 
a minimal separation
and then they start to depart from each other.
This observation reveals, thereby, 
a repulsive character of the inter-hump interaction.
So, during the many-body dynamics the interaction between the humps of the initially-prepared
$0$-phase symmetric wave-packet
changes from attractive to repulsive and stays repulsive afterwards.

The many-body dynamics of the initially-prepared $\pi$-phase anti-symmetric wave-packet
is shown on the right middle panel, Fig.~1e.
Due to the larger separation between the two humps,
the spatial density profiles obtained at the many-body and GP levels 
are indeed very close to each other.
The weak repulsive interaction between the humps 
persists at the many-body level as well.
The right lower panel, Fig.~1f, 
shows the density profile 
for the $0$-phase asymmetric wave-packet constructed,
as above, from two functions of slightly different widths.
The dynamics obtained at the many-body level for 
the initially-prepared $0$-phase asymmetric wave-packet
has a lot in common with the respective many-body dynamics obtained for the 
initially-prepared $0$-phase 
symmetric
wave-packet,
compare panel Fig.~1d and Fig.~1f.
The sub-clouds start to attract each other and approach some minimal distance, 
and then the inter-hump interaction becomes of repulsive character.
Summarizing, the many-body dynamics shows that irrespective to the 
phase between initially-prepared humps,
after some time the inter-hump interaction changes its nature and becomes repulsive.
This conclusion is general and has been checked for various initially-prepared symmetric 
and asymmetric wave-packets with different phases. 
This behavior is in accord with the experimental observation that 
the inter-hump interactions in low-dimensional attractive Bose gases are 
indeed repulsive. 
For more analysis of the 
inter-hump forces see \cite{SM}.

The differences between the GP dynamics leading to bright soliton trains
and the respective many-body dynamics collected in Fig.~1 show that 
bright soliton trains in one-dimensional BECs undergo changes in time. 
We shall see below that these changes are fundamental and dramatic. 
To get a deeper insight into the physics behind the intriguing many-body results, 
we examine further the quantum nature of the attractive bosons. 
First, we perform the natural orbital analysis of the propagating wave-packets 
at every point of the propagation time by constructing and diagonalizing
the reduced one-body density matrix.
The reduced one-body density matrix of the system is 
given by $\rho^{(1)}(x|x';t) = 
\langle\Psi(t)|\hat{\mathbf \Psi}^\dag(x')\hat{\mathbf \Psi}(x)|\Psi(t)\rangle$,
where $\hat{\mathbf \Psi}^\dag(x)$ is the usual
bosonic field operator creating a boson at position $x$.

The eigenvalues obtained, called natural occupation
numbers, 
are plotted in Fig.~2.
All initial states, at $t=0$, are fully condensed systems, 
i.e., only one natural orbital is macroscopically occupied.
The GP theory implies that the system remains condensed all the time, i.e.,
that only one natural orbital remains occupied during the evolution. 
The many-body dynamics, in contrast, 
shows that both for the initially-prepared 
$0$-phase soliton train 
as well as for the $\pi$-phase soliton train 
the bosons quickly
populate the second natural orbital,
forming thereby macroscopically (two-fold) 
fragmented many-boson states.
The fate of bright soliton trains is thus death by fragmentation.
We note that fragmentation of BECs is a general known and well-studied phenomena in the literature
(see \cite{frag1,frag2,frag3} and references therein).
For the $0$-phase symmetric initial state the fragmentation ratio of $n_1 = 63.21\%$, 
where the condensed fraction has decreased by $1 - \frac{1}{e}$ times, 
is reached already at $t \approx 30$
(see left upper panel of Fig.~2).
Afterwards, the many-body wave-packet dynamics reveals oscillations around the two-fold fragmented state.

The $\pi$-phase soliton loses its coherence monotonously
and lives a bit longer -- it becomes 
fragmented at $t \approx 54$ (see right upper panel of Fig.~2).
We stress that the changes in the physics are
dramatic despite the large
separation between the density humps  
(see Fig.~1e).
The lifetime of the $0$-phase asymmetric soliton (not shown)
discussed above and in Fig.~1 is also $t \approx 30$.
We have also found out that 
increasing the number of atoms in the 
system and keeping the GP parameter $\lambda_0 N$ fixed, 
the lifetime of the soliton trains decreases, i.e., 
the lifetime of the initially-prepared coherent objects becomes even shorter \cite{SM}.
The conclusions are that initially coherent two-hump wave-packets 
become two-fold fragmented with time, 
and that the resulting coherence lifetime of the 
two-hump solitons is finite and ultrashort.

We have performed similar many-body propagations 
of initially coherent three-hump and four-hump soliton trains 
and found that the dynamics leads to the formation of three-fold 
and four-fold fragmented many-body states, 
respectively. 
Summarizing, the death of multi-hump soliton trains is a generic 
rapid inevitable feature of the many-body dynamics of attractive bosons 
in one dimension, 
and gives birth to a potentially interesting object 
like the fragmenton \cite{fragmenton},
also see \cite{SM}. 
This result has been obtained for initially 
coherent states made of different numbers of atoms, 
different number of the constituting humps, 
placed at different inter-hump separations, 
with different inter-hump phases, 
and of different humps' widths.

We have shown so far the death of soliton trains as 
an emerging many-body phenomenon in one dimensional 
attractive BECs.
It is important to discuss how can it be resolved experimentally.
Looking at Fig.~1 we recall that on the many-body
level the multi-boson attractive wave-packets
remain localized objects in real 
space.
This observation, 
together with the fragmentation of the 
reduced one-body density matrix, 
suggest to look for signatures in the correlation 
functions, 
both in coordinate and momentum space,
which is done in Fig.~3.
The first-order correlation function
in coordinate space $g^{(1)}(x',x;t) \equiv \frac{\rho^{(1)}(x|x';t)}{\sqrt{{\rho(x,t) }{\rho(x',t)}}}$
quantifies the degree of spatial coherence of the interacting system \cite{Glauber,Kaspar_corr},
and analogously $g^{(1)}(k',k;t)$ in momentum space \cite{Kaspar_corr}.
The upper two panels
show the first-order correlation functions
in coordinate (left) and momentum (right)
space of the initially-prepared coherent
$0$-phase symmetric soliton train at $t=0$.
It should be stressed that, 
within GP theory, 
the wave-packet remains coherent at all times;
thus the upper two panels represent the
first-order correlation functions of the
two-hump attractive wave-packet at all times, 
assuming GP to be valid.
In sharp contrast,
the first-order correlation functions
in coordinate and momentum space,
when computed on the many-body level, 
show prominent structures, 
see middle and lower four panels in Fig.~3, 
which leave no place for confusion of
the rise of fragmentation \cite{Kaspar_corr}
and the death of soliton trains.
Another quantity of interest
is the momentum distribution
of the soliton train itself,
shown in \cite{SM}.

It is left to determine whether the death of soliton
trains is on a time scale relevant for current experimental setups. 
For this we 
consider explicitly $N=2000$ $^7$Li atoms ($m = 1.1650 \cdot 10^{-26}$ kg, 
$s$-wave scattering length is $a_s = -3a_0$). 
We emulate a quasi-1D cigar-shaped trap in which the transverse confinement is $w_\perp = 2\pi \cdot 800$ Hz,
which is amenable to current experimental setups.
Following \cite{olsh}, the transverse confinement renormalizes the interaction strength.
Combining all the above,
the length scale is given by 
$L = \frac{\hbar |\lambda_0|}{2 m \omega_\perp a_s} = 11.3 \cdot 10^{-6}$ m,
and the time scale by $\frac{m L^2}{\hbar} = 14.2 \cdot 10^{-3}$ sec.
Now,
expressing the lifetime of the soliton trains (see Fig.~2) in real time,
we get lifetimes of a couple of hundreds of milliseconds (see \cite{SM} for 
analysis of the lifetime as a fucntion of $N$), 
much below current experimental 
run times of a couple of seconds \cite{exp1}.
Our theoretical results thus call
for exciting experimental demonstration of
the ultrafast death of soliton trains
in low-dimensional attractive BECs,
caused by
fragmentation.

We are grateful to 
K. Sakmann and H.-D. Meyer for fruitful discussions.
Financial support by the DFG is greatly acknowledged.

\newpage
\thispagestyle{empty}

\begin{figure}[]
\vglue -1.5 truecm
    \centering
    \includegraphics[width=12cm,angle=0]{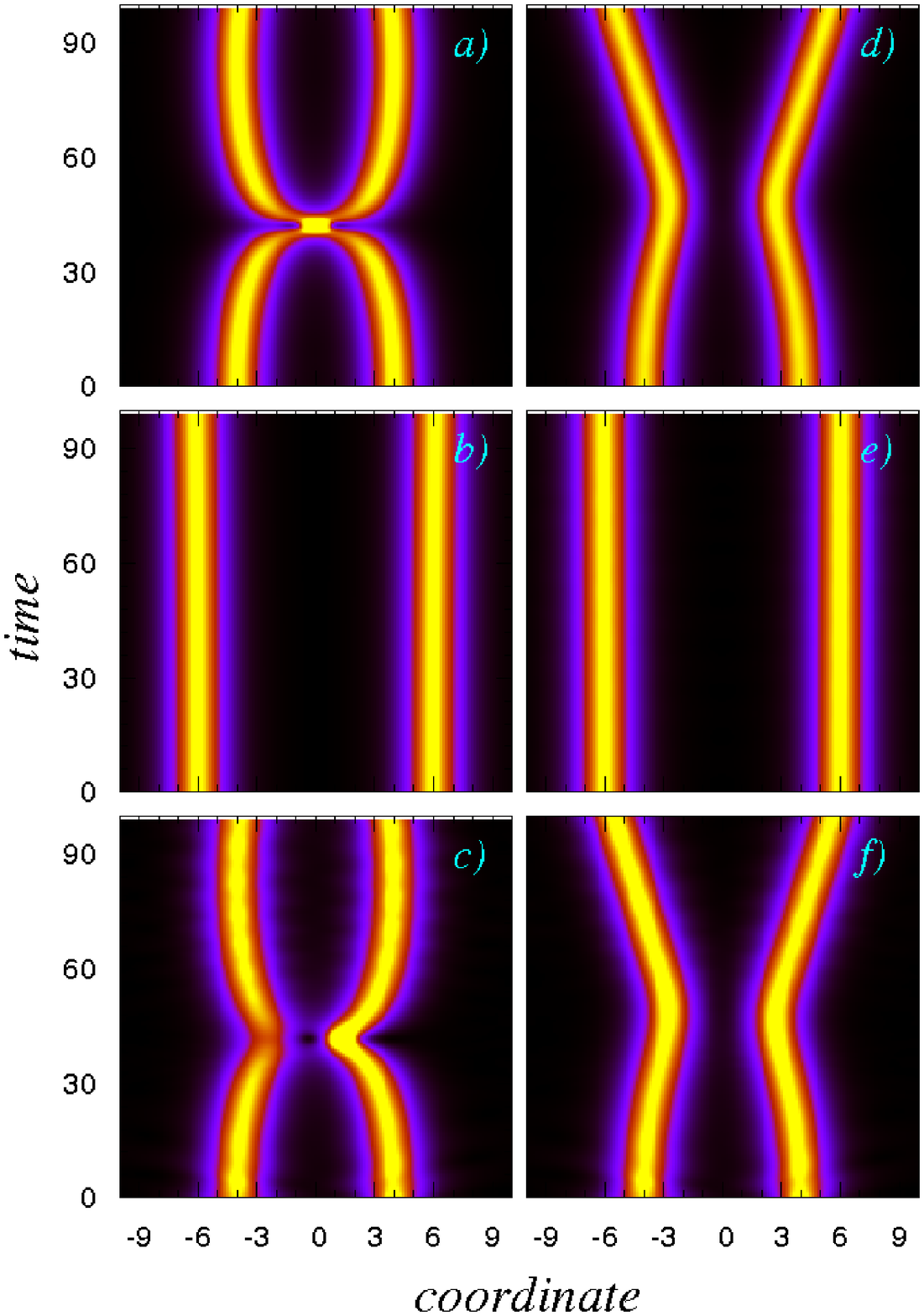}
    	\caption{(color online). Evolution in time of a two-hump soliton train in coordinate space.
Shown is the evolution in time of the density $\rho(x,t)$ of the attractive system 
with $N=2000$ bosons and $\lambda_0=-0.002$ computed 
by the popular Gross-Pitaevskii equation on the mean-field level in panels a)-c),
and the corresponding dynamics computed on the many-body level in panels d)-f).
Panels a) and d) show the dynamics of a $0$-phase symmetric train;
Panels b) and e) show the dynamics of a $\pi$-phase anti-symmetric train;
Panels c) and f) show the dynamics of a $0$-phase asymmetric train.
On the many-body level the forces between
the two density humps are weaker, 
and there is essentially only repulsion and no attraction 
between them at all but short times.
All quantities shown are dimensionless.}
    \label{fig1}
\end{figure}

\begin{figure}[]
    \centering
    \includegraphics[width=12cm,angle=-90]{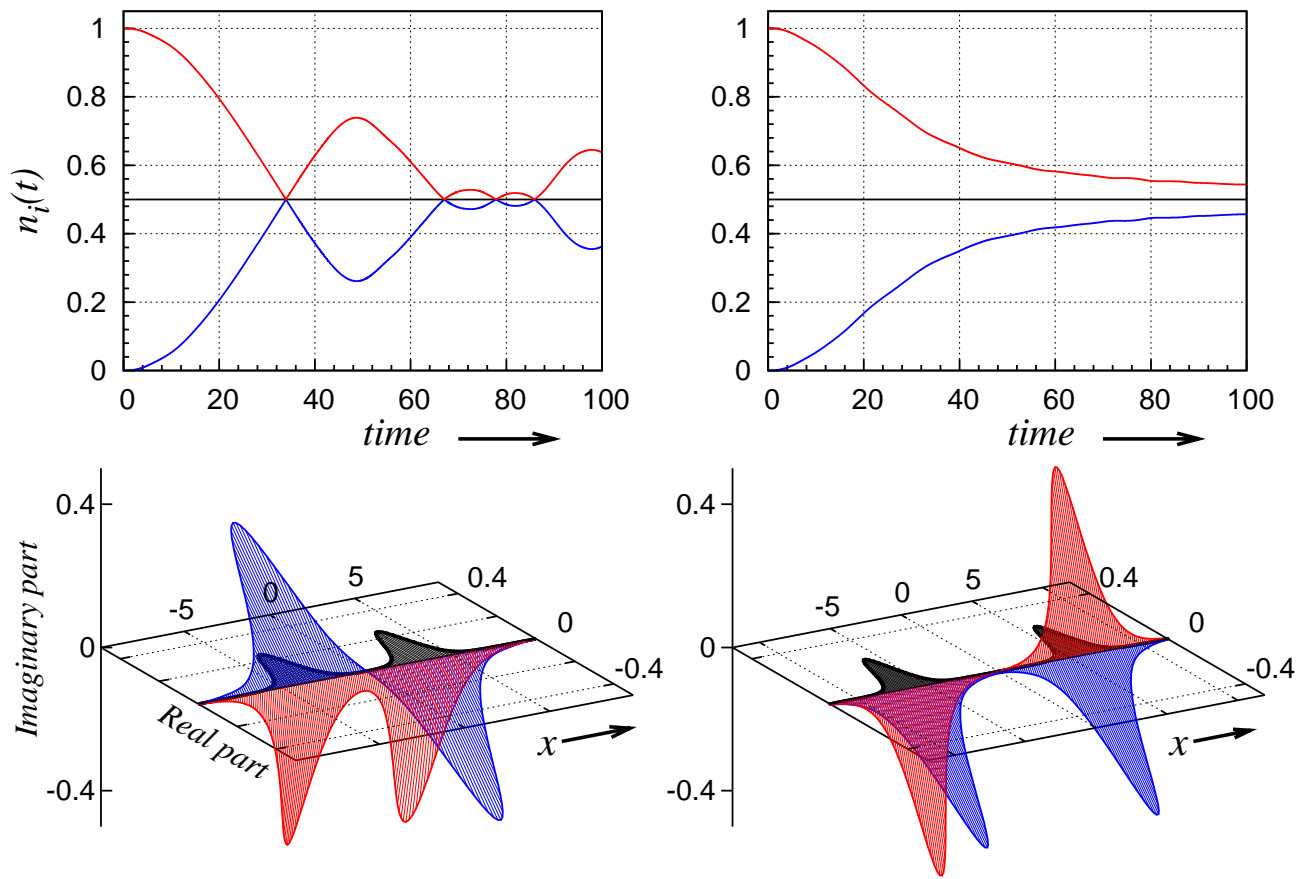}
    	\caption{(color online). Death of soliton trains by fragmentation.
The upper panels show the occupation numbers of 
the reduced one-body density matrix
of the initially-prepared 
$0$-phase symmetric (left) and $\pi$-phase anti-symmetric (right) 
soliton trains of Fig.~1.
It is seen that the initially condensed systems become
quickly fragmented.
The emergence of fragmentation signifies
the death of the condensed objects known as bright soliton trains.
The lower two panels depict the respective natural orbitals 
after fragmentation
of the 
initially-prepared $0$-phase symmetric (left) and $\pi$-phase 
anti-symmetric (right) coherent wave-packets.
The two natural orbitals are delocalized 
functions of 'gerade' (even) and 'ungerade' (odd) symmetries.
The delocalization of the natural orbitals in combination with
the exchange interaction stabilizes the
fragmented attractive Bose-Einstein condensates
in comparison with the condensed soliton trains.
All quantities shown are dimensionless.}
    \label{fig2}
\end{figure}

\begin{figure}[]
\vglue -2.0 truecm
    \centering
    \includegraphics[width=12cm,angle=0]{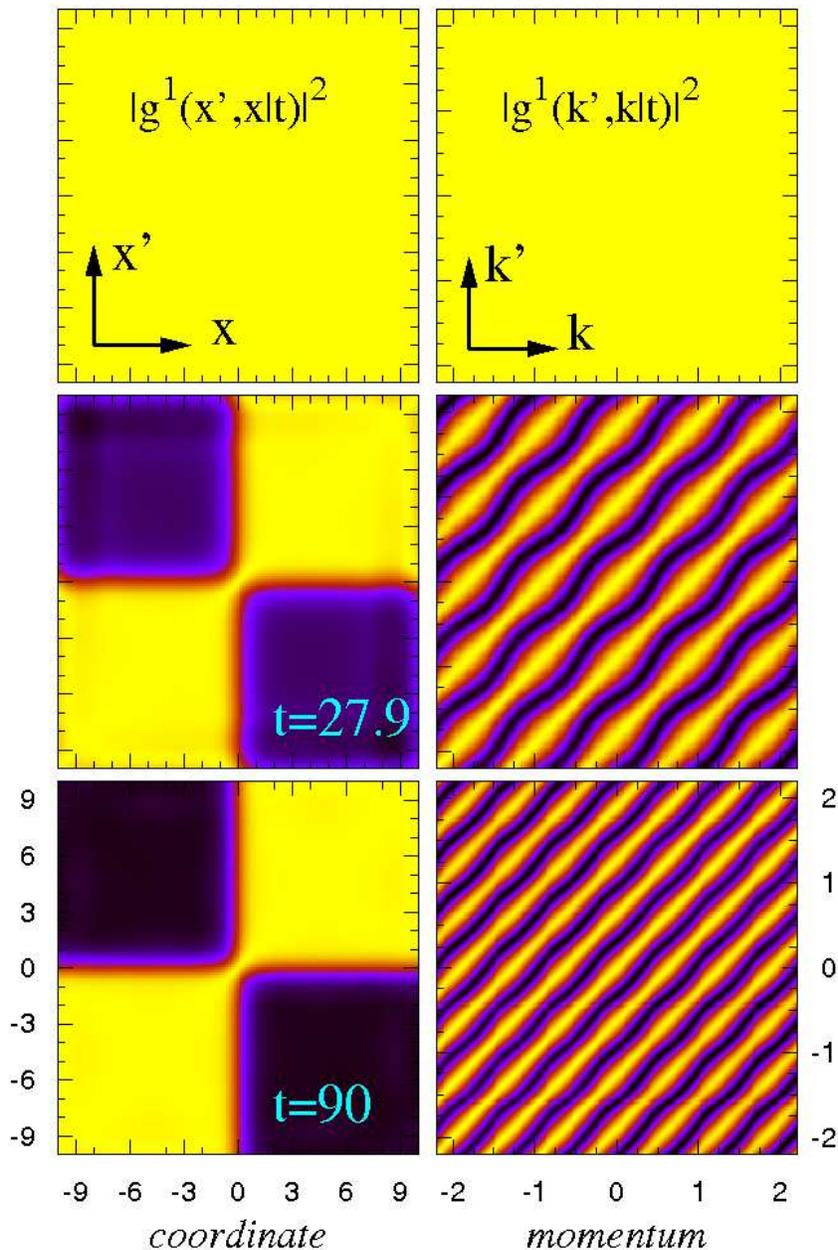}
        \caption{(color online). Detecting the death of soliton trains by means of first-order correlation functions.
The upper two panels show the first-order correlation function
of the initially-prepared 
$0$-phase symmetric soliton train 
of Fig.~1 in coordinate (left) and momentum (right) space.
It is completely flat signifying a coherent 
and condensed state.
Within Gross-Pitaevskii theory the system remains 
condensed for all times; thus the upper two panels represent
the first-order correlation function in this case for all times.
On the many-body level, the correlation functions
are completely different and structured.
The pattern emerging in coordinate space indicates
no coherence between the two humps (left middle and lower panels),
the respective pattern in momentum space 
represents a fragmented state (right middle and lower panels).
Thus, measuring the first-order correlation functions
in either coordinate or momentum space
provides a direct indication of 
the death of the soliton train.
All quantities shown are dimensionless.}
    \label{fig3}
\end{figure}

%%%%%%%%%%%%%%%%%%%%%%%%%%%%%%%%%%%%%%%
\newpage
\clearpage

\begin{center}
\Large{\bf Supplemental Material} 
\end{center}

\section*{Forces between density humps in attractive objects}

The time-dependent densities
of the evolving initially-prepared bright soliton trains 
shown in Fig.~1 reveal intricate "interactions"
between the density humps.
On the Gross-Pitaevskii (GP) or non-linear Schr\"odinger equation 
level (see in this respect Refs.~[S1,S2]),
$0$-phase symmetric and asymmetric trains show 
attraction between the density humps,
which manifests
itself as 
"bound" periodic-like motion of the density humps
one with respect to the other.
The $\pi$-phase anti-symmetric train
reveals repulsive interaction 
between the density humps,
which would eventually lead to their 
increasing separation 
from each other. 

On the many-body level,
we have seen in Fig.~1 that repulsive interaction
between the density humps eventually prevails, 
even when the $0$-phase symmetric and asymmetric trains 
are used as the initially-prepared wave-packets of the attractive system. 
In this section we would like to contract a simple mean-field model to
qualitatively explain the forces between the density humps 
when soliton trains become fragmented and die.

Let us commence with the development of fragmentation in the $0$-phase symmetric system.
The natural orbitals of the decaying soliton train (see left lower panel of Fig.~2)
are composed of 'gerade' (even) and 'ungerade' (odd) delocalized functions,
i.e., they simultaneously exist in the right and left sub-clouds.
Oversimplifying the many-body picture of the decay of soliton trains
and the development of fragmentation in the system, 
we can imagine that at every point in time the system is 
described by a two-orbital mean-field state made
of a single permanent (Fock state)
$|n_1,n_2\rangle$ [S3]
with time-dependent occupation numbers and orbitals.
The orbitals comprising this permanent 
are the above described delocalized even and odd functions.
In time,
the initially-prepared condensed state $|N,0\rangle$
(built from the delocalized even orbital only, see left panels of Fig.~2) 
becomes fully fragmented at about $t \approx 34$  
(see left upper panel of Fig.~2).
Then it is described by the mean-field state
$|N/2,N/2\rangle$
(built from both the delocalized even and odd orbitals).
Such a fragmetned state has been recently
predicted to exist in one dimensional attractive BECs
and called fragmenton [S4].
What we would like now to compare
within this simplified model is
the inter-hump forces in the condensed state $|N,0\rangle$
and fully fragmented state $|N/2,N/2\rangle$.
For completeness,
the inter-hump forces in the condensed state $|0,N\rangle$
(built from the delocalized odd orbital only, see right panels of Fig.~2) 
will be registered as well.
The delocalized even and odd functions used in our model
are conveniently taken as the $0$-phase symmetric
and $\pi$-phase anti-symmetric linear combinations
${\cal N}\!\left\{\mathrm {sech} [\gamma(x-x_0)] \pm \mathrm {sech} [\gamma(x+x_0)]\right\}$,
which reproduce on the mean-field level
the limiting cases of the bright soliton trains 
and the fragmenton.

The energy of the Fock state $|n_1,n_2\rangle$ 
is given as the expectation value of the Hamiltonian
$E(\gamma,x_0,n_1,n_2) = \langle n_1,n_2|\hat H|n_1,n_2\rangle$
and depends on the parameters $\gamma$ (inversely proportional to the width of
the density humps) and $x_0$ (half the separation between the density humps),
and on the relative occupation numbers $n_1$ and $n_2$ ($n_1 + n_2 = N$).
By taking the derivative $ - \frac{\partial E}{\partial x_0}$ one immediately arrives at the inter-hump force $F(\gamma,x_0,n_1,n_2)$.
The full expression is quite lengthy, but in the large $x_0\gamma>2$ limit,
i.e., when the humps are well separated
from each other,
we can prescribe an asymptotic analytical force in the lowest order:
$$
F(\gamma,x_0,n_1,n_2)  = - \frac{\partial E (\gamma,x_0,n_1,n_2)}{\partial x_0} 
$$
$$
= - \frac{4 e^{-2 \gamma x_0}\gamma^2}{3N}(n_1-n_2) \left[\lambda_0 (N-1)(\gamma x_0 -2) + 4 \gamma^2 x_0-5 \gamma \right].
 \eqno{\mathrm {(S1)}}
$$
For pure condensed states, i.e., when $n_1=N, n_2=0$ or $n_1=0,n_2=N$ with the optimal width $\gamma_{GP}=\lambda_0 (N-1)/4$,
this general expression takes on a very simple form (see in this respect Ref.~[S2]):
$$
F(\gamma_{GP},x_0,N)= - \frac{\partial E (\gamma_{GP},x_0,N)}{\partial x_0}
= \mp 4 e^{-2 \gamma_{GP} x_0}\gamma_{GP}^3, \eqno{\mathrm {(S2)}}
$$
where the "-" corresponds to 
the $0$-phase symmetric and the "+"
to the $\pi$-phase anti-symmetric two-hump soliton.
The damping exponent originates from the residual overlap 
between the left and right sub-clouds.

In Fig.~S1 we plot the full inter-hump forces and their asymptotic expansions given in Eqs.~(S1,S2) 
as a function of the inter-hump separation $2x_0$ for
the system of $N=2000$ bosons and
$\lambda_0=-0.002$ discussed in the main text.
In the one limiting case,
$n_1=N, n_2=0$, 
the force between the density humps
of the $0$-phase symmetric soliton train is negative, i.e., attractive.
In the other limit, $n_1=0, n_2=N$,
the force between the density
humps of the $\pi$-phase anti-symmetric soliton train is positive, i.e., repulsive.
For the pure fragmenton $|N/2,N/2\rangle$ this force vanishes asymptotically, 
because of the $(n_1-n_2)$ prefactor in Eq.~(S1).
This fact allows us to predict that the inter-hump interaction in the fragmented states is always weaker than in the respective
two-hump coherent soliton trains.
Interestingly, 
the residual interaction for the 
pure fragmenton at a large but not yet asymptotic separation is, 
nevertheless, repulsive.
By going from one limit to another, i.e., by transferring the bosons, say, from 
the $0$-phase symmetric 
to the $\pi$-phase anti-symmetric
natural orbital, 
we see that the character of the interaction gradually changes from attractive to repulsive.

\section*{Lifetime of two-hump matter-wave soliton trains}

In the present section we would like to investigate
the scaling of the lifetime of two-hump solition trains with the number
of bosons $N$ in the system.
We concentrate on the $0$-phase symmetric soliton train.
To make the comparison,
we keep the GP parameter $\lambda_0 N = -4$ 
and inter-hump separation $2x_0 = 8$ fixed
to the values used for $N=2000$ throughout the main text.
It should be reminded that for a fixed value of  $\lambda_0 N$
the GP dynamics is independent of $N$ (for $N \gg 1$, which is the situation here).
Additionally, no fragmentation can be described by the GP equation, of course,
and solitons' lifetime is infinite.
Hence, the many-body results presented 
in this section are to further demonstrate the
substantial physical differences between
the mean-field and many-body quantum dynamics 
of 1D attractive Bose systems.

Fig.~S2 collects the results for the lifetime as a function of $N$.
We remind that the lifetime is defined as the time $t$ where 
the fragmentation ratio of $n_1 = 100\% \times \left(1 - \frac{1}{e}\right) = 63.21 \%$ is reached.
Translating from dimensionless to dimension-full 
quantities, 
the lifetime of, e.g., the two-hump soliton made of 
$N=2000$ $^7$Li atoms
in a quasi-1D cigar-shaped trap with transverse confinement of $w_\perp = 2\pi \cdot 800$ Hz
is about $400$ milliseconds.
This lifetime is short compared to 
current experimental setups
which are a couple of seconds long.
It is seen from Fig.~S2 
that
the lifetime decreases roughly quadratically 
with the number of bosons $N$.
For $N=8000$ $^7$Li atoms it is below $50$ milliseconds.
All in all,
we demonstrate that the
death of soliton trains is an ultrafast process that takes place
on a time scale which is an order of magnitude
shorter than relevant time-scales for 
current experimental setups.

\section*{Momentum distribution of initially-prepared two-hump soliton trains}

The density profile of bright soliton trains consists of localized structures.
As we have seen in the main text this well-known GP (mean-field)
characteristic persists on the many-body level as well,
despite the fact that the system becomes fragmented.
Consequently, 
the spatial density of low-dimensional attractive systems
is not a straightforward 
indicator for the development of fragmentation in the system,
and hence is not likely to be utilized experimentally to identify it.
In this section we would like to augment the indicting 
information on the death of solitons by reporting 
the corresponding momentum distribution, see Fig.~S3.

The time-dependent momentum distribution $n(k,t)$
obtained by solving the time-dependent GP equation for
the two-hump solitons of Fig.~1 is plotted in Fig.~S3.
In the left upper panel,
Fig.~S3a, we plot the momentum distribution
of the $0$-phase symmetric soliton train.
A sharp contrast and oscillatory pattern is seen,
which signifies the system being condensed 
(and coherent, see right upper panel of Fig.~3).
As the two density humps
approach each other in real space,
the corresponding momentum
distribution broadens.
At the collision point ($t \approx 42$)
the momentum distribution
is the broadest,
since the two colliding clouds momentarily 
merge into one cloud,
yet its maximal value 
is significantly smaller.

In the left middle panel, Fig.~S3b,
the momentum distribution of the
$\pi$-phase anti-symmetric soliton is depicted. 
Due to the larger separation between the density 
humps ($x_0 = \pm 6$, see Fig.~1b),
more maxima and narrower peaks in $n(k,t)$ are seen, 
in comparison to the soliton train of Fig.~S3a
(for which $x_0 = \pm 4$,  see Fig.~1a).
Furthermore, 
the momentum distribution is essentially
constant for all times shown,
as the corresponding spatial density is,
due to the initial larger separation 
between the density humps. 
Finally, in Fig.~S3c,
the momentum distribution
of the $0$-phase asymmetric  
soliton train is shown.
It also depicts a sharp contrast pattern, 
but asymmetric.
The broadest peaks in $n(k,t)$ 
correspond
to the near collision of
the two density humps 
in real space occuring at about $t \approx 42$
(see Fig.~1c).

We may summarize the following features of the 
momentum distribution of two-hump solitons
computed by the GP theory: (i) appearance of sharp contrast pattern
signifying condensation (and coherence), and
(ii) the number of peaks seen and their width 
depend on the initial separation of the density humps 
in real space (in a reminiscence of a double-slit 
experiment with light).
   
On the many-body level, the
time-dependent momentum distribution $n(k,t)$
shows essential visual differences from the GP momentum distribution,
compare the left and right panels of Fig.~S3.
In the right upper panel,
Fig.~S3d,
the many-body momentum-distribution 
dynamics of the initially-prepared $0$-phase
symmetric soliton train is shown.
For times up to about $t \approx 15$,
the sharp contrast pattern characterizing the
initially-prepared condensed system survives.
Looking at the evolution of the corresponding occupation 
numbers (see left upper panel of Fig.~2),
we see that up to this time the
system remains at least $90\%$ condensed.
For times longer than just about $t \approx 15$,
the many-body momentum distribution almost suddenly becomes
"blared", in sharp contrast to the GP results.
This is yet another signature of the
fragmentation of the system.
The "blared" momentum distribution 
pattern can be anticipated by "interfering"
the momentum 
distributions carrying the two 
delocalized even and odd natural orbitals 
of the fragmented system (see in this respect the 
left lower panel of Fig.~2).
Similarly, the many-body momentum distribution of
the initially-prepared $\pi$-phase anti-symmetric
wave-packet, Fig.~S3e,
and the many-body 
momentum distribution
of the initially-prepared $0$-phase asymmetric 
wave-packet, Fig.~S3f, quickly become "blared",
when fragmentation sets it.
We stress that,
despite the large
separation between the density humps  
of the $\pi$-phase anti-symmetric
wave-packet (see Fig.~1e),
the changes in the physics are dramatic.

Summarizing,
the momentum distribution
of low-dimensional attractive systems 
initially-prepared as (condensed and coherent)
soliton trains quickly loose their contrast and become
essentially single-peaked and "blared".
This consequence and fingerprint of fragmentation 
should be amenable to experimental detection,
and is far better and straightforward indicator 
of fragmentation
than the monitoring of localization in coordinate space.

\makeatletter
\renewcommand*{\@biblabel}[1]{[S#1]}
\makeatother

\newpage
\thispagestyle{empty}

\addtocounter{figure}{-3}

\renewcommand{\figurename}{Figure S\hglue -0.12 truecm}

\begin{figure}[]
    \centering
    \includegraphics[width=12cm,angle=-90]{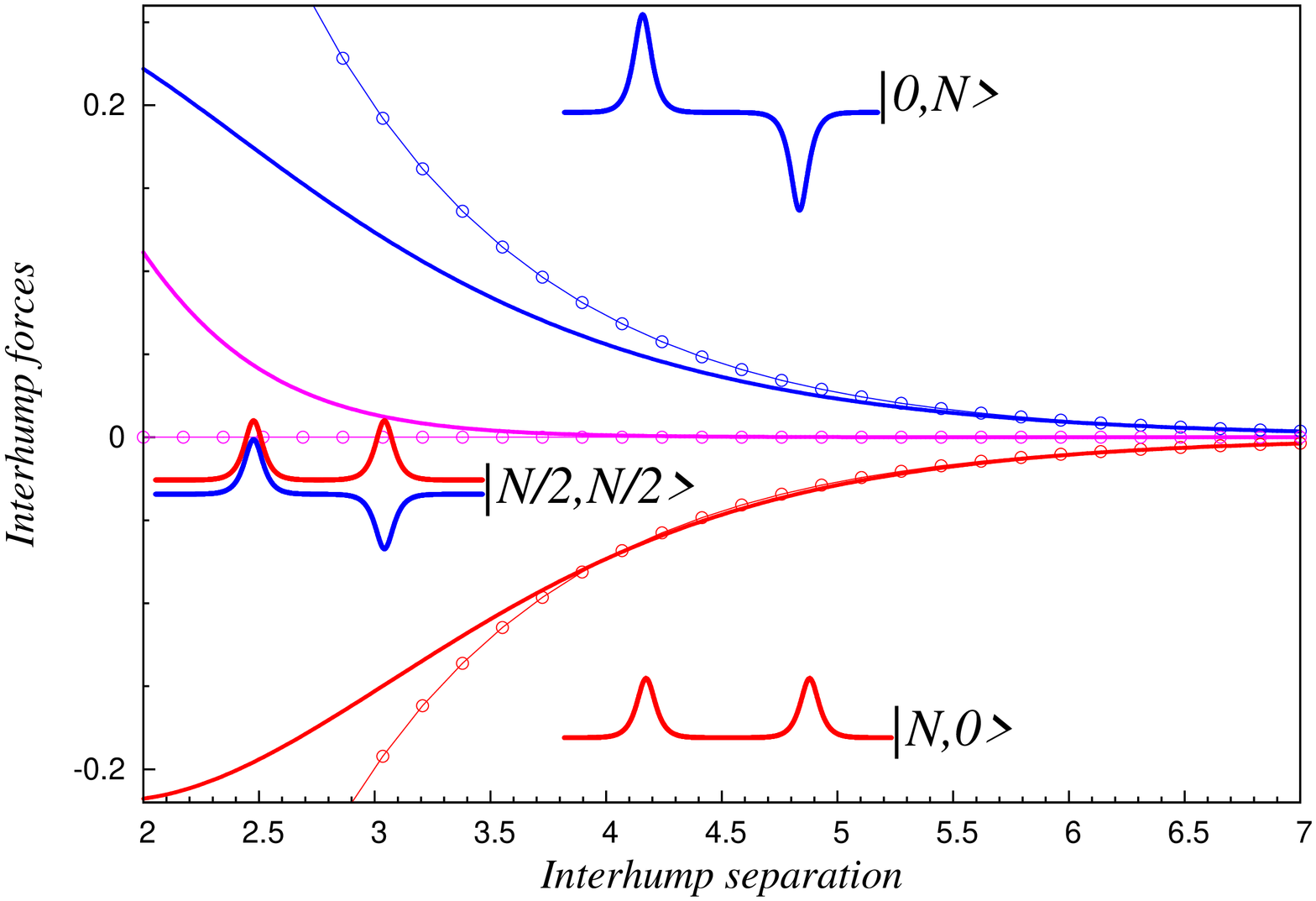}
    	\caption{(color online). Forces between density humps in a soliton train and fragmenton.
Shown are the forces as a function of inter-hump separation $2x_0$
for the $0$-phase symmetric soliton train (attractive force)
and the $\pi$-phase anti-symmetric soliton train (repulsive force)
for the system of $N=2000$ bosons with $\lambda_0=-0.002$.
The inter-hump force in a fragmented state built
as an admixture of the two soliton trains, 
$0$-phase symmetric and $\pi$-phase anti-symmetric 
two-hump solitons (the fragmenton),
is much weaker than for the soliton trains themselves,
and is repulsive.
The full-line curves depict
the forces computed numerically,
whereas the circle-line curves the
asymptotic forces given in Eq.~(S1) for the fragmenton
and Eq.~(S2) for the soliton 
trains. All quantities shown are dimensionless.}
    \label{figS1}
\end{figure}

\begin{figure}[]
    \centering
    \includegraphics[width=12cm,angle=0]{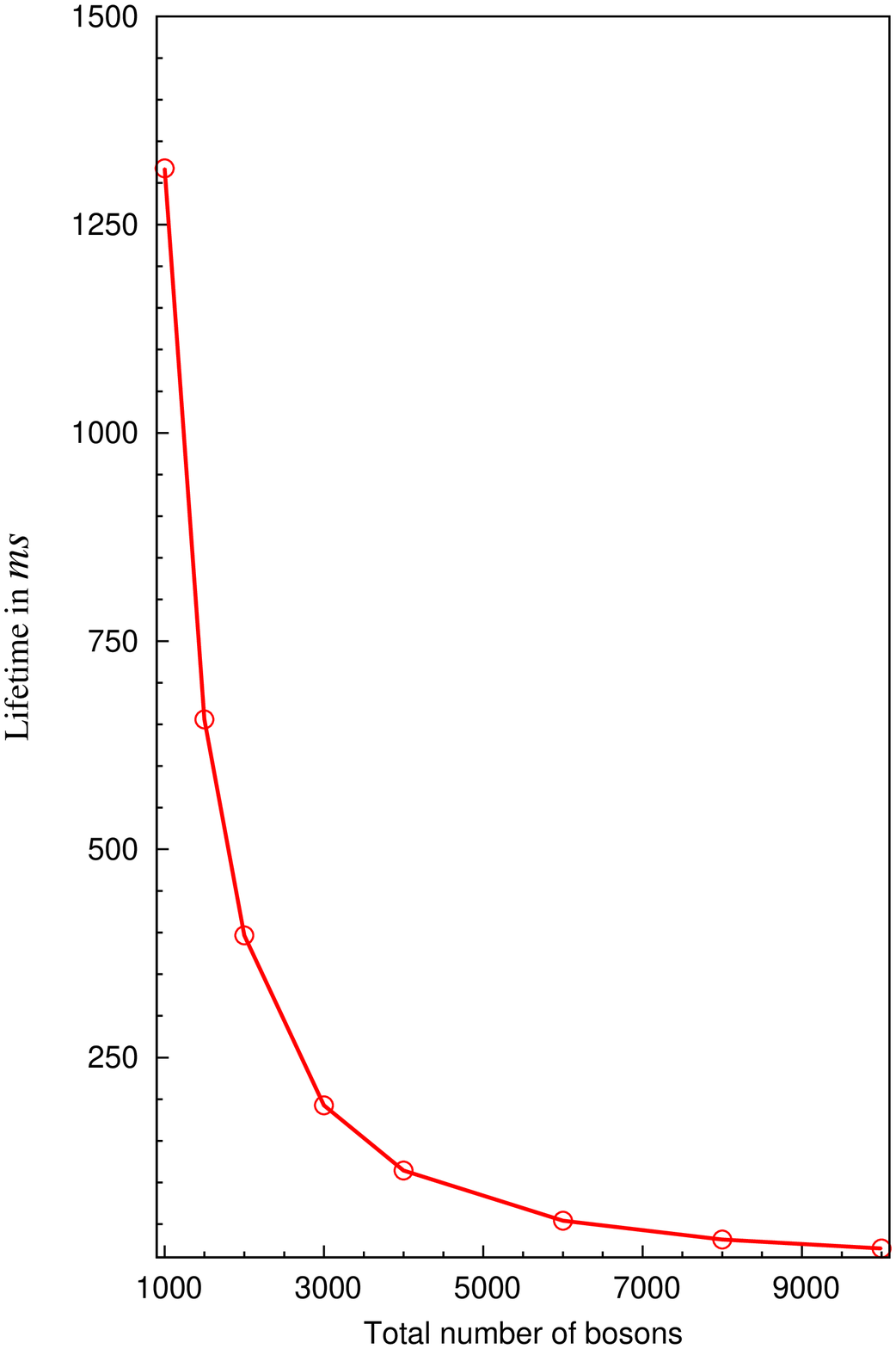}
    	\caption{(color online). Lifetime of two-hump soliton trains.
Depicted is the lifetime in milliseconds of the $0$-phase symmetric 
two-hump soliton trains
as a function of the total number of bosons $N$
(the solid line connecting the open circles
is to guide the eye only). 
The Gross-Pitaevskii parameter $\lambda_0 N = - 4$ and inter-hump
separation $2x_0 = 8$ are kept fixed 
(corresponding to the system of $N=2000$ bosons with
$\lambda_0=-0.002$
discussed in the main text). 
The lifetime decreases roughly quadratically 
with the number of bosons $N$.
It is demonstrated that the
death of soliton trains is an ultrafast process that can take place
on a time scale which is an order of magnitude
shorter than relevant time-scales for 
current experimental setups
(couple of seconds).}
    \label{figS2}
\end{figure}

\begin{figure}[]
\vglue -2.0 truecm
    \centering
    \includegraphics[width=12cm,angle=0]{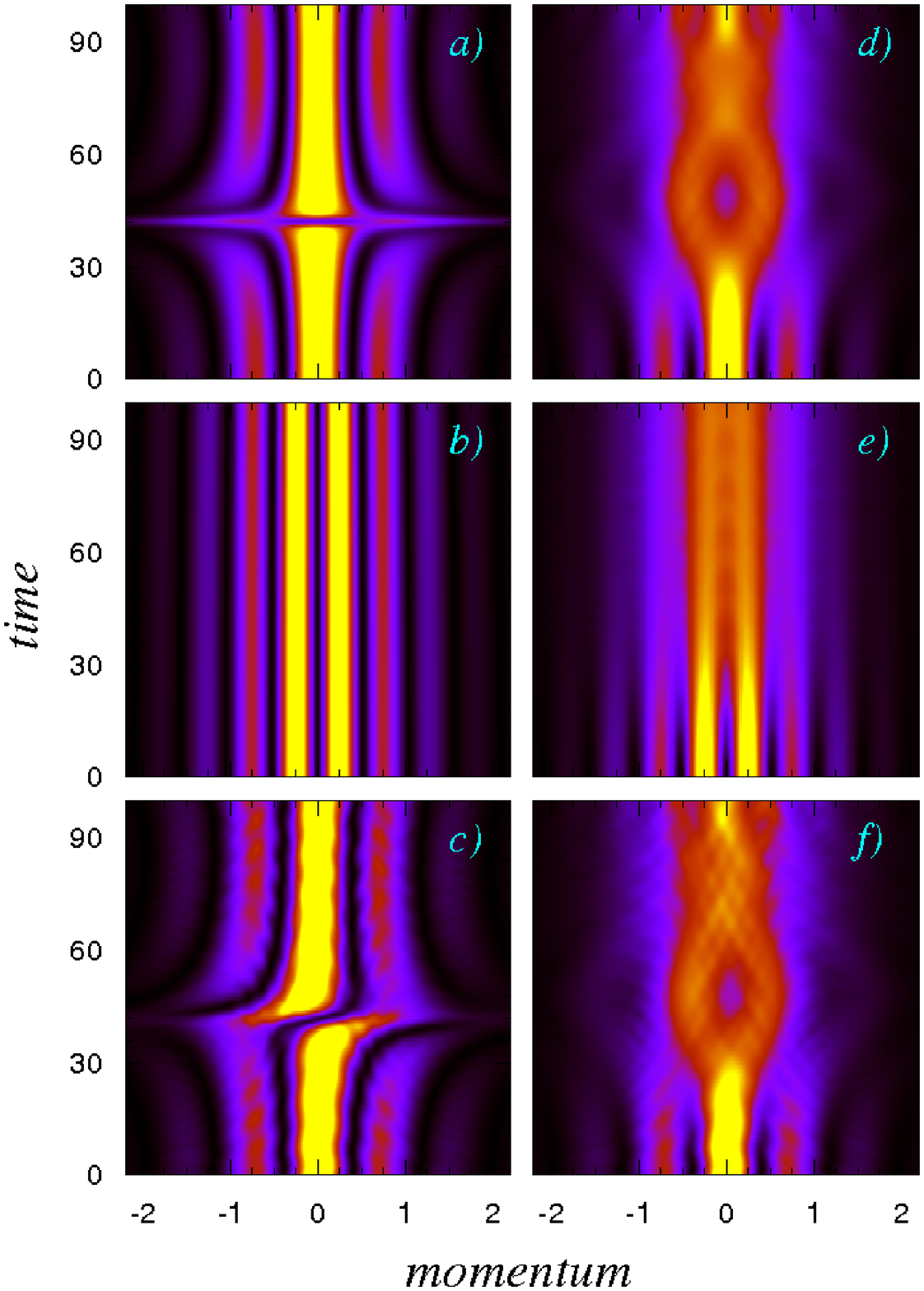}
        \caption{(color online). Evolution in time of a two-hump soliton train in momentum space.
Shown is the evolution in time of the momentum distribution $n(k,t)$ of the attractive system 
of $N=2000$ bosons with $\lambda_0=-0.002$ computed 
by the popular Gross-Pitaevskii equation on the mean-field level in panels a)-c),
and the corresponding dynamics computed on the many-body level in panels d)-f).
Panels a) and d) show the dynamics of a $0$-phase symmetric train;
Panels b) and e) show the dynamics of a $\pi$-phase anti-symmetric train;
Panels c) and f) show the dynamics of a $0$-phase asymmetric train.
On the mean-field level, 
panels a)-c), an interference-like sharp-contrast pattern emerges,
indicating the two-hump system being condensed.
On the many-body level, panels d)-f), 
the momentum distribution is much more blared,
indicating the fragmentation of the system.
All quantities shown are dimensionless.}
    \label{figS3}
\end{figure}


\begin{thebibliography}{64}

\bibitem{sol1} S. Trillo and W. Torruellas (Eds.),
{\it Spatial Solitons} 
(Springer, Berlin, 2001).

\bibitem{sol2}  G. I. Stegeman and M. Segev, 
Science {\bf 286}, 1518 (1999).

\bibitem{solth1}  P. A. Ruprecht, M. J. Holland, K. Burnett, M. Edwards,
Phys. Rev. A {\bf 51}, 4704 (1995).   

\bibitem{solth2}  V. M. P\'erez-Garc\'ia, H. Michinel, and H. Herrero,     
Phys. Rev. A {\bf 57}, 3837 (1998).

\bibitem{exp1}  K. E. Strecker, G. B. Partridge, A. G. Truscott, and R. G. Hulet,
Nature {\bf 417}, 150 (2002).

\bibitem{exp2} L. Khaykovich {\it et al.},
Science {\bf 296}, 1290 (2002).

\bibitem{attrac1} U. Al Khawaja {\it et al.},
Phys. Rev. Lett. {\bf 89}, 200404 (2002).

\bibitem{attrac2} L. Salasnich, A. Parola, and L. Reatto,  
Phys. Rev. Lett. {\bf 91}, 080405 (2003).

\bibitem{attrac3} L. D. Carr and J. Brand,  
Phys. Rev. Lett. {\bf 92}, 040401 (2004).

\bibitem{attrac4} A. D. Martin, C. S. Adams, and S. A. Gardiner, 
Phys. Rev. Lett. {\bf 98}, 020402 (2007).

\bibitem{GP1} E. P. Gross, 
Nuovo Cimento {\bf 20}, 454 (1961).

\bibitem{GP2}  L. P. Pitaevskii,
Zh. Eksp. Teor. Fiz. {\bf 40}, 646 (1961)
 [Sov. Phys.--JETP {\bf 13}, 451 (1961)].

\bibitem{frag1} P. Nozi\`eres and D. Saint James,
J. Phys. France {\bf 43}, 1133 (1982). 

\bibitem{frag2} A. I. Streltsov, O. E. Alon, and L. S. Cederbaum,
Phys. Rev. A {\bf 73}, 063626 (2006).

\bibitem{frag3}  E. J. Mueller,  T.-L. Ho, M. Ueda, and G. Baym,
Phys. Rev. A {\bf 74}, 033612 (2006).   

\bibitem{MCTDHB_lett}  A. I. Streltsov, O. E. Alon, and L. S. Cederbaum,
Phys. Rev. Lett. {\bf 99}, 030402 (2007).

\bibitem{MCTDHB_pap} O. E. Alon, A. I. Streltsov, and L. S. Cederbaum,
Phys. Rev. A {\bf 77}, 033613 (2008).

\bibitem{Grond} J. Grond, J. Schmiedmayer, and U. Hohenester,  
Phys. Rev. A {\bf 79}, 021603(R) (2009).

\bibitem{BJJ_us} K. Sakmann, A. I. Streltsov, O. E. Alon, and L. S. Cederbaum, 
Phys. Rev. Lett. {\bf 103}, 220601 (2009).

\bibitem{SM} See EPAPS Document No. xxxx for 
supplemental physical analysis of the 
death of soliton trains.
For more information on EPAPS, see http://www.aip.org/pubservs/epaps.html.

\bibitem{fragmenton} A. I. Streltsov, O. E. Alon, and L. S. Cederbaum,
Phys. Rev. Lett. {\bf  100}, 130401 (2008).

\bibitem{Glauber} M. Naraschewski and R. J. Glauber,
Phys. Rev. A {\bf 59}, 4595 (1999).

\bibitem{Kaspar_corr} K. Sakmann, A. I. Streltsov, O. E. Alon, and L. S. Cederbaum,
Phys. Rev. A {\bf  78}, 023615 (2008).

\bibitem{olsh} M. Olshanii,  
Phys. Rev. Lett. {\bf 81}, 938 (1998).




\end{thebibliography}

\begin{thebibliography}{10}

\bibitem{sol2_Segev} G. I. Stegeman and M. Segev, 
Science {\bf 286}, 1518 (1999).

\bibitem{sol_force}  V. I. Karpman and V. V. Solov'ev,
Physica D {\bf 3}, 487 (1981).  

\bibitem{BMF}  A. I. Streltsov and L. S. Cederbaum,
Phys. Lett. A {\bf  318}, 564 (2003).

\bibitem{fragmenton_str} A. I. Streltsov, O. E. Alon, and L. S. Cederbaum,
Phys. Rev. Lett. {\bf  100}, 130401 (2008).

\end{thebibliography}
\end{document}